 \documentclass{elsart}
 \usepackage{epsfig}
\begin{document}

\begin{frontmatter}
  \title{
  ELECTRON SPECTROSCOPY \\
  AND DENSITY-FUNCTIONAL STUDY \\
  OF ``FERRIC WHEEL'' MOLECULES}
 \author[aut1]{
 A.V. Postnikov,\corauthref{cor}\thanksref{label1}}
 \author[aut1]{S.G. Chiuzb\u{a}ian,}
 \author[aut1]{M. Neumann,}
 \author[aut1,aut2]{S. Bl\"ugel}
 \address[aut1]{
 University of Osnabr\"uck -- Department of Physics,
  D-49069 Osnabr\"uck, Germany}
 \address[aut2]{
 Institut f\"ur Festk\"orperforschung,
 Forschungszentrum J\"ulich,
 D-52425 J\"ulich, Germany}
\thanks[label1]{Permanent address: Institute of Metal Physics,
                620219 Yekaterinburg, Russia}

 \corauth[cor]{Corresponding Author: Tel. +49-541-9692377;
  Fax +49-541-9692351; E-mail: \underline{apostnik@uos.de}}

\begin{abstract}
The Li-centered ``ferric wheel'' molecules with six oxo-bridged
iron atoms form molecular crystals. We probed their electronic
structure by X-ray photoelectron (XPS) and soft
X-ray emission spectroscopy (XES), having calculated in
parallel the electronic structure of a single ``ferric wheel''
molecule from first-principles by tools of the density-functional
theory, using, specifically, the {\sc Siesta} method. The Fe local
moments were found to be 4 $\mu_{\mbox{\tiny B}}$, irrespective
of their mutual orientation. Neighbouring atoms, primarily oxygen,
exhibit a noticeable magnetic polarization, yielding effective
spin $S$=5/2 per iron atom, that can get inverted as a ``rigid''
one in magnetic transitions. Corresponding energy preferences can
be mapped onto the Heisenberg model with effective exchange
parameter $J$ of about $-$80 K.
\end{abstract}

\end{frontmatter}

\section{Introduction}
\label{sec:intro}

A family of ring-form metal-organic complexes of iron
\cite{AnChIE36-2482} has attracted an appreciable
interest due to their natural beauty and high symmetry, favourable
for building theoretical models. Waldmann \emph{et al.}
\cite{InCh38-5879} analyzed the data of magnetic susceptibility
measurements, torque magnetometry and inelastic neutron
scattering, yielding a satisfactory mapping onto the spin
Hamiltonian
$$
H = -J \left(\sum_{i=1}^5 {\bf S}_i{\cdot}{\bf S}_{i+1}
+ {\bf S}_6{\cdot}{\bf S}_1 \right) +
(\mbox{anisotropy term}) + (\mbox{Zeeman term})\,.
$$
with $J\sim$ 18 -- 20 K (depending on sample and method)
for \{Li$\subset$Fe$_6$[N(CH$_2$CH$_2$O)$_3$]$_6$\}Cl,
the system of our present study. The
ground state was found to be antiferromagnetic; the charge state
of iron in such materials is routinely referred to as Fe(III),
that implies individual spins $S$=5/2 in the above equation.
While there is recently a consensus reached in interpreting
magnetic bulk measurements in terms of the Heisenberg spin model
\cite{PRL89-246401,PRB60-1161,InCh40-2986}, the microscopic
information on the electronic subsystem is much less known.
We undertook a number of X-ray spectroscopy
measurements, of which we report now the photoelectron
valence band spectrum and the Fe $L_3$ X-ray emission spectrum.
These results are compared with, and discussed on the basis of,
first-principles electronic structure calculations carried out in the
framework of the density-functional theory (DFT). This gives an
additional insight in the site-resolved magnetization, chemical
bonding, and energetics of magnetic transitions in ``ferric
wheels''.

\section{X-ray photoelectron and emission spectra}
\label{sec:spec}

The XPS measurements were performed using a PHI Model 5600ci
MultiTechnique system and employing monochromatized Al~$K_{\alpha}$
radiation. The pressure in the vacuum chamber during the
measurements was below 5${\cdot}$10$^{-9}$~mbar; the charge
neutrality on the surface was achieved by a low energy electron
flood gun. The energy resolution as determined at the Fermi level
of a gold foil was 0.3--0.4~eV. Powder samples 
of \{Li$\subset$Fe$_6$[N(CH$_2$CH$_2$O)$_3$]$_6$\}Cl, obtained by drying
the crystals in vacuum and thus removing the CHCl$_3$ molecules,
were measured in a small gold crucible. The structure of the
hexagonal unit cells is preserved upon drying, but the lattice
parameters $a$ and $b$ are reduced by about 30\%
\cite{InCh38-5879}. The calibration of the XPS spectra was
performed assuming two non--equivalent positions of C in the
molecule and 0.8~eV difference between the corresponding C$1s$
peaks, as suggested by the similar surrounded C atoms in PTMG and
PEI polymers (see Ref.~\cite{Scienta_DB}).
With this procedure the maximum of the C$1s$
XPS spectrum was found at 286~eV. The XES measurements were
performed at Beamline 8.0.1 at the Advanced Light Source of
the Lawrence Berkeley National Laboratory. The spectra were
taken employing the University of Tennessee at Knoxville soft
X-ray fluorescence endstation \cite{RSI66-1394}. The Fe~$L_{3,2}$ emission 
was measured with incident photon energies ranging between 702.2 and
738.7~eV and an energy resolution of about 0.8~eV. The spectra
were calibrated using a pure Fe sample as reference. In this
report only the spectrum corresponding to the Fe~$L_3$ emission is
listed.

\section{Electronic structure calculations}
\label{sec:calcul}

The calculations of electronic structure from first-principles
have been done within the DFT, using the
{\sc Siesta} method \cite{JPCM14-2745}. The features of this
method are an efficient construction of localized
numerical atom-centered basis functions and the use of
norm-conserving pseudopotentials. The independence on crystal
symmetry makes the method particularly useful for structure
optimizations and molecular dynamics. In the present case,
however, because the calculations on such large system are
quite time-consuming, we performed them so far only for nominal (fixed) 
structures.

``Ferric wheels'' crystallize along with some amount of solvent
molecules; due to the fact that intermolecular distances are large,
we included in the present simulation only a single molecule
of 140 atoms -- without solvent, but including a chlorine atom
(actually present in molecular crystal) as a counter-ion to lithium,
for correct charge compensation. The molecule was treated in a box
with dimensions 22$\times$22$\times$18 {\AA}, on which the real-space 
grid of 180$\times$180$\times$150 divisions (corresponding to
the cutoff energy of 180 Ry) was imposed for solving the Poisson equation
by the fast Fourier transform. The basis set included
double-$\zeta$ functions with polarization orbitals on O and Fe and
double-$\zeta$ functions on all other atoms (according to the specifications
used by the tight-binding community; see for details Ref.~\cite{JPCM14-2745}
and references therein).

The calculations have been performed using the generalized gradient
approximation of Perdew, Burke and Ernzerhof \cite{PRL77-3865}.
Although non-collinear magnetization density can be treated by {\sc Siesta},
for a large system as the present one this slows down calculations
considerably. As the ground state of our ``ferric wheel'' is believed
to be antiferromagnetic (AFM), it is simulated as an alternating sequence
of up and down spin moments on the ring. Other spin arrangements -- 
fully ferromagnetic (FM), or with one or two spins inverted -- have also 
been tried, and found stable. The energy relations between these solutions 
are discussed in the next section.

\begin{figure}
\centerline{\epsfig{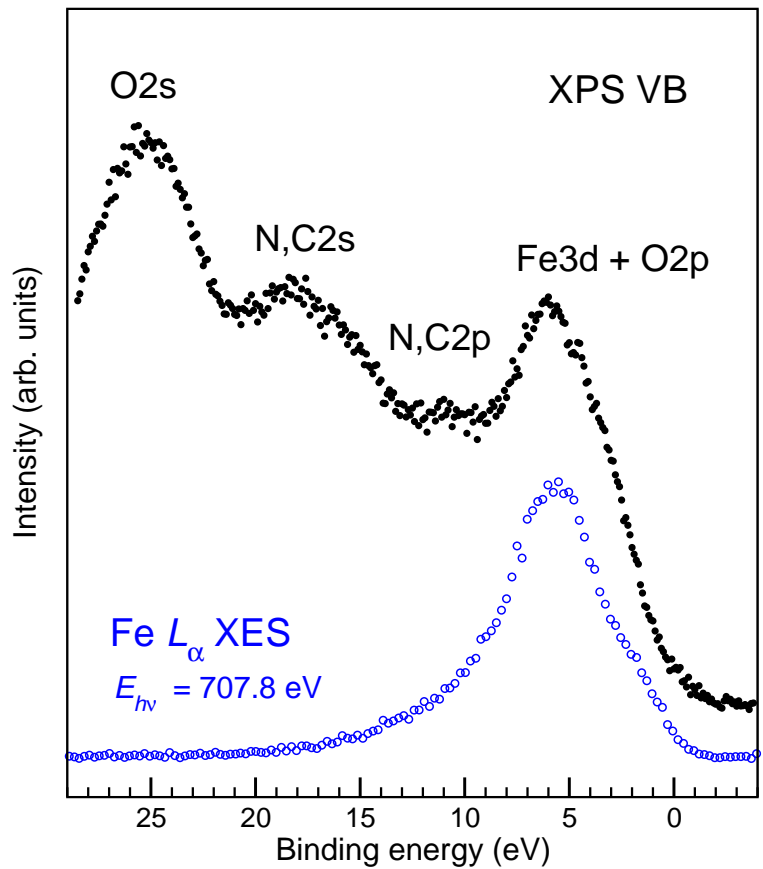}}
\bigskip
\caption{\label{fig:Spec}
Measured XPS spectrum of the valence band and
Fe $L_3$ X-ray emission spectrum, obtained with
the excitation energy 707.8 eV.
\bigskip
}
\end{figure}

\begin{figure}
\centerline{\epsfig{file=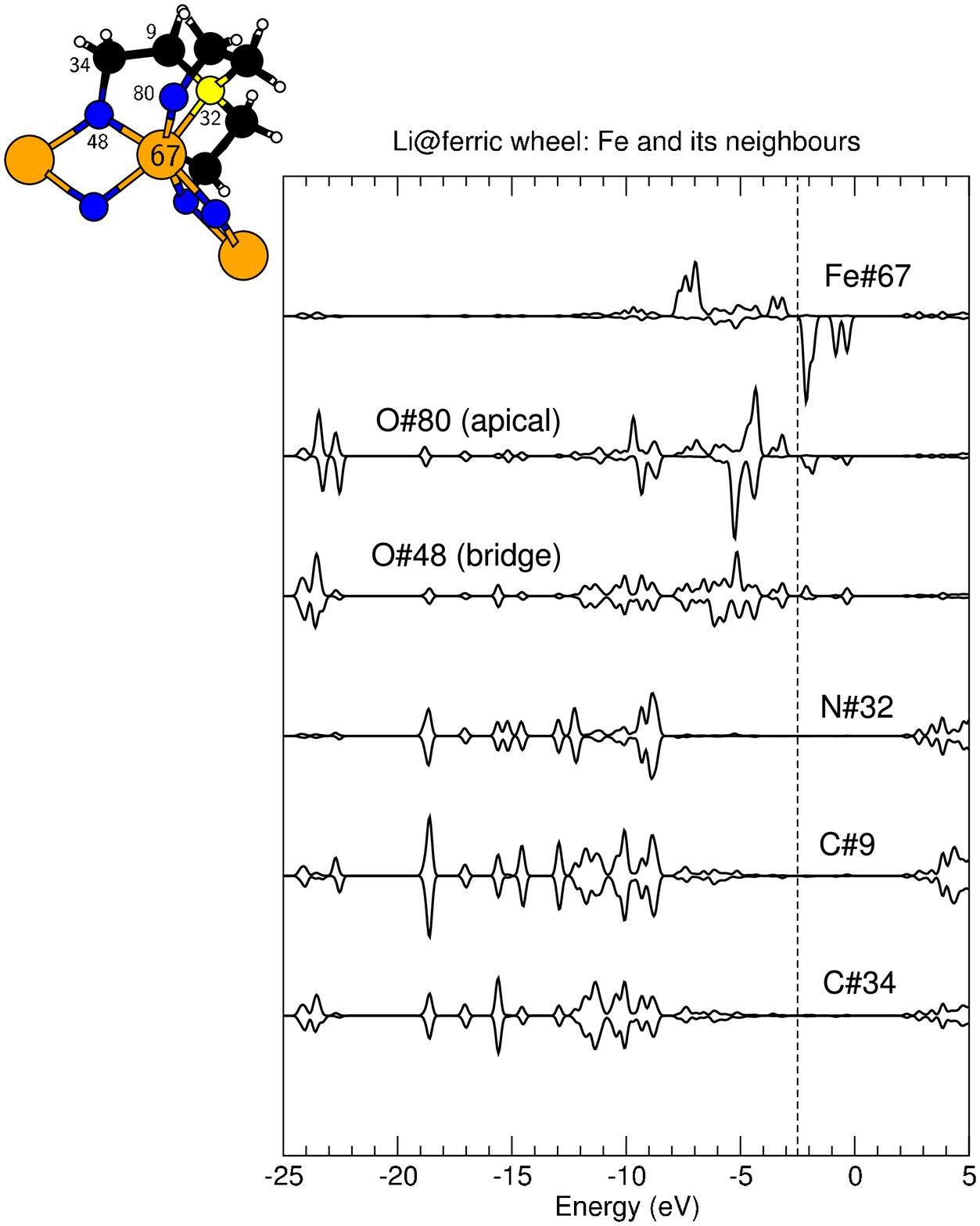,width=16.0cm}}
\bigskip
\caption{\label{fig:PDOS}
Partial DOS at the Fe atom and its several neighbours
in the ``ferric wheel'' molecule (as labeled in the inset).
The Fermi energy is indicated by a dashed vertical line.
\bigskip
}
\end{figure}

\section{Results and discussion}
\label{sec:results} 
The photoelectron spectrum of
the valence band reveals separated features which can be
easily attributed to O, N and C $2s$ states, and a roughly
structureless peak around the binding energy of $\sim$5 eV, that
can be tentatively attributed to Fe$3d$ states, hybridizing with
the O$2p$ band (see Fig.~\ref{fig:Spec}). A comparison with
electronic structure results confirms this assignment. The Fe
$L_3$ X-ray emission spectrum probes occupied Fe$3d4s$ states; it
reveals an overall shape in good agreement with calculated partial
density of Fe$3d4s$ states. The latter turns out to be very
similar irrespective of the actual mutual
orientation of spins, FM or antiferromagnetic
AFM, and exhibits in all cases a crystal-field splitting (of
about 4 eV) into $t_{2g}$-like and $e_g$-like states in the
majority-spin channel. The calculated partial DOS at the Fe site,
along with some its neighbours, is shown in Fig.~\ref{fig:PDOS}.

These results reveal a complicated pattern of hybridization
between Fe, O and N atoms, a detailed discussion of which will be
given elsewhere. One notes a pronounced Fe$3d$--O$2p$ chemical
bonding, which gives rise to magnetic polarization in oxygen $2p$ shells.
Moreover, more distant carbon and nitrogen atoms, whose $2p$
states are moreover quite separated in energy from the
Fe$3d$, exhibit a small but not vanishing polarization.

\begin{figure}
\hspace*{-1.0cm}{\epsfig{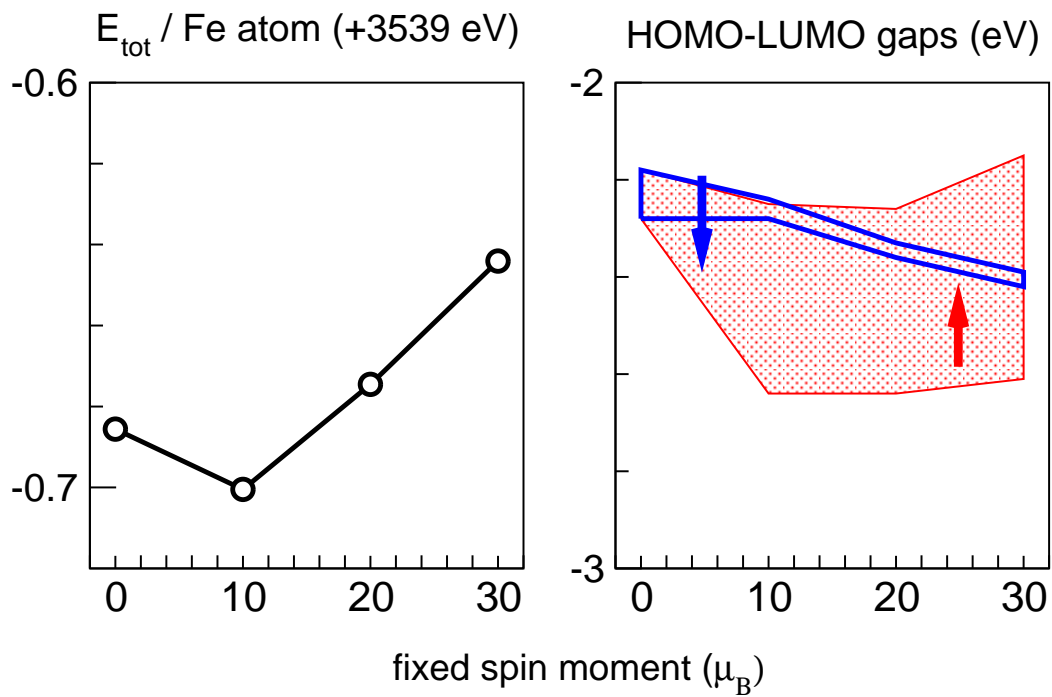}}
\bigskip
\caption{\label{fig:FSM}
Total energy per Fe atom (left panel)
and energy gap in two spin channels
(right panel; shaded area -- majority-spin,
thick lines -- minority-spin) from fixed spin moment calculations.
\bigskip
}
\end{figure}

It is noteworthy that the $3d$-shell of Fe, although
strongly polarized, does not yet develop a spin configuration
$S$=5/2, contrary to what is routinely assumed in the literature.
Instead, the local magnetic moment is close to 4 $\mu_{\mbox{\tiny
B}}$, in relative independence of its precise definition (choice
of functions for the population analysis, etc.). While the
majority-spin states are fully occupied, the minority-spin band
still contains about one electron, because of the above
mentioned Fe$3d$--O$2p$ hybridization. The presence of one
minority-spin electron on the Fe site implies the Fe$^{2+}$
configuration and not Fe$^{3+}$ as would seem compatible with $S$=5/2. 
The peculiarity of the situation is that an extra majority-spin
electron is delocalized over the organic fragment hosting
the Fe atom. This recovers spin $S$=5/2 per Fe atom in the
molecule, consistently with macroscopic measurements.

The partial DOS shown in Fig.~\ref{fig:PDOS} corresponds to some
artificial broadening of discrete energy levels of the molecule,
for better visibility. Such broadening is legitimate, as long as
it remains small enough not to smear out the band gap and to maintain a clear
separation between groups of energy levels split in the ligand field.
Actually, an artificial broadening is a useful tool for speeding up
the convergence of calculation, that (due to the presence of many
nearly degenerate levels near the Fermi energy) is otherwise
quite instable even at mixing parameters of about 10$^{-4}$,
demanding for many thousands of iterations. The problem is that
such smearing does affect calculated total energies quite dramatically,
to the point that a reliable comparison of spin-flip energetics, used
to estimate exchange interactions of the order of tens of K, becomes
impossible. A very useful technique in this relation is the fixed spin moment
(FSM) scheme \cite{JPF14-L129}.  In our case, imposing an (integer) 
spin moment per molecule fixes the number of electrons in both spin channels 
and thus removes the major reason of instability in the course of calculations.
As a consequence, the broadening parameter can be safely reduced
to the values which permit a reliable extrapolation to zero level width.
The results conveniently converge for FSM values of 30 $\mu_{\mbox{\tiny B}}$
(FM case), 20 and 10 $\mu_{\mbox{\tiny B}}$ (one and two spins inverted,
correspondingly); 0 (alternate-spin AFM case). The calculated total energy
values are shown in Fig.~\ref{fig:FSM}.

Typically, comparison of total energies obtained in the FSM
scheme needs to take into account the ``Zeeman term'' due to the Fermi
energies -- now possibly different in two spin channels -- moving apart,
as if in a fictitiously imposed magnetic field. In our case, however,
no additional terms to the total energy must be included, because --
as is seen in the right panel of Fig.~\ref{fig:FSM} -- there is a common
energy gap in both spin channels throughout the whole range of FSM values
we studied. A linear change of the total energy while inverting one and then
two spins from the FM configuration implies the validity of the Heisenberg
model, as we seemingly deal with ``rigid'' magnetic moments. An additional
argumentation comes from the fact that the magnitudes of local magnetic
moments always remain close (within several per cent)
to 4 $\mu_{\mbox{\tiny B}}$, and the partial DOS on Fe sites remains largely
unaffected by the actual magnetic ordering. Keeping this in mind, and
assuming Heisenberg-model spin Hamiltonian as in Section~\ref{sec:intro}
with the $S$ value of 5/2 (i.e., for the total spin which gets inverted),
we arrive at the estimate for $-J$ of around 80 K (over both 30$\rightarrow$20
and 20$\rightarrow$10 $\mu_{\mbox{\tiny B}}$ flips). This is qualitatively
correct (i.e. indicates preferentially AFM coupling) and even of correct
order of magnitude. However, two observations can be done.
First, the ``true'' AFM configuration (with half of magnetic moments
inverted on the ring) does not follow the linear trend (see Fig.~\ref{fig:FSM})
and lies actually higher in energy than the configuration with two spins
inverted. The origin of this is not yet clear to us at the moment.
{}From one side, the zero-FSM configuration is -- technically -- the most
difficult to converge, so some numerical instability can still play a role.
{}From the other side, the true AFM ground-state configuration is apparently
not collinear, the fact which does not follow from the simplest
(nearest-neighbours) version of the Heisenberg model and probably manifests
its shortcomings.
The second observation concerns the magnitude of exchange parameter
$J$ and the fact that it is probably overestimated by a factor of $\sim$4
in our calculation. The origin of this lies most probably in
on-site correlations, which, if treated accurately beyond the standard
schemes of the DFT, would primarily affect localized Fe$3d$ states,
shifting the bulk of occupied states downwards in energy, the bulk of
unoccupied states upwards, expanding the energy gap, and -- whatever scheme
to use for estimating exchange parameters -- substantially reducing their
magnitude. This has been recently shown for another molecular magnet
(Mn$_{12}$) by Boukhvalov \emph{et al.} \cite{PRB65-184435}.
\footnote{We note, that the LDA+$U$ formalism \cite{LDA+U} is
a possible scheme of choice for such a simulation. It may help to adjust
the situation in a physically reasonable way, but does not necessarily help
to make quantitative \emph{predictions}.}

Summarizing our analysis of the electronic structure of a Li-based
``ferric wheel'', the combined use of spectroscopy techniques and
first-principles calculations leads to the following conclusions:
This system develops strong local magnetic moments on Fe sites,
but their magnitude is close to 4 $\mu_{\mbox{\tiny B}}$
and not 5 $\mu_{\mbox{\tiny B}}$ as is often assumed.
The value of 5 $\mu_{\mbox{\tiny B}}$ can still be attributed to
a Fe atom \emph{in a molecule}, with considerable part of
this magnetization distributed over (mostly) oxygen and other
atoms. Correspondingly, iron atoms are not ionized to +3, but
instead acquire a configuration close to Fe$^{2+}$,
with a substantial covalent part in the Fe--O bonding.
With respect to its magnetic interactions, this system can be mapped
reasonably well onto the Heisenberg model; hence we deal
with \emph{rigid} magnetic moments which are nevertheless
\emph{delocalized} -- an interesting counter-example
to a common belief that the Heisenberg model primarily applies
to localized spins.

\section*{Acknowledgements}
The authors thank the Deutsche Forschungsgemeinschaft for
financial support (Priority Program ``Molecular Magnetism'').
A.V.P. acknowledges useful discussion with Jens Kortus and
J\"urgen Schnack. The samples for measurements and crystal
structure data have been kindly provided by the group of Prof.
Dr. R.~W.~Saalfrank from the Institute for Organic Chemistry of the
University Erlangen-N{\"u}rnberg.
The work in the Advanced Light Source at the Lawrence Berkeley 
National Laboratory was supported by the U.S. Department of Energy 
(Contract No. DE-AC03-76SF00098).


\end{document}